\documentclass{sf2a-conf2013}
\usepackage{graphicx}
\usepackage{hyperref}
\usepackage[]{natbib}  
\usepackage{epstopdf}

\def\BibTeX{{\rm B\kern-.05em{\sc i\kern-.025em b}\kern-.08em
    T\kern-.1667em\lower.7ex\hbox{E}\kern-.125emX}}
\bibpunct{(}{)}{;}{a}{}{,}  

\begin{document}

\TitreGlobal{SF2A 2013}


\title{Modelling the relative velocities of isolated pairs of galaxies}

\runningtitle{The model relative velocities of isolated pairs of galaxies}

\author{V. Gonzalez-Perez}\address{Centre de Physique des Particules de Marseille, Aix-Marseille Universit\'e,  CNRS/IN2P3, Marseille, France.}

\author{E. Jennings}\address{The Kavli Institute for Cosmological Physics, University of Chicago, 5640 South Ellis Avenue, Chicago, IL 60637, USA; The Enrico Fermi Institute, University of Chicago, 5640 South Ellis Avenue, Chicago, IL 60637, USA}

\author{M.-C. Cousinou$^1$}

\author{S. Escoffier$^1$}

\author{A. Tilquin$^1$}

\author{A. Ealet$^1$}

\setcounter{page}{237}


\maketitle


\begin{abstract}
We study the comoving relative velocities, $v_{12}$, of model isolated galaxy pairs 
at $z=0.5$. For this purpose, we use the predictions from the {\sc galform} semi-analytical model 
of galaxy formation and evolution based on a $\Lambda$ cold dark matter 
cosmology consistent with the results from WMAP7. In real space, we find that isolated pairs of galaxies are predicted to form an angle $t$ with the line-of-sight that is uniformly distributed as expected if the Universe is homogeneous and isotropic. We also find that isolated 
pairs of galaxies separated by a comoving distance between 1 and 3 $h^{-1}$Mpc 
are predicted to have $\left<v_{12}\right>\sim 0$. For galaxies in this regime, the distribution of the angle $t$ is predicted to change minimally from real to redshift space, with a change smaller than 5\% in $\left< {\rm sin}^2t\right>$. However, the distances defining the {\it comoving regime} strongly depends on the applied isolation criteria. 
\end{abstract}



\section{Introduction}

The observed expansion of the Universe is attributed to a dark energy component
 of which little is known \citep[e.g.][]{blake11,anderson12}. \citet{mb} (MB) 
derived a geometrical test on isolated pairs of galaxies that can provide an independent investigation of the abundance and nature of the dark energy. The idea  of this test is to measure the relative angle, $t$, that isolated pairs of galaxies form with the line-of-sight (LOS). The pairs of galaxies can be thought off as dumbbells. In a homogeneous and isotropic universe, the orientation of these dumbbells in the sky will be uniformly distributed and thus, will have a probability distribution of the form sin $t/2$. Thus, measuring the probability distribution of the angle $t$ can put constraints on the cosmological parameters, in particular, on the characteristics of the dark energy \citep{ap79,phillipps94}. However, galaxies are affected by local gravitational pulls which are separate to the effect of dark energy and MB developed a test taking these into account. Here, we study the average relative comoving velocities, $\left<v_{12}\right>$, of pairs of galaxies
 at $z=0.5$ drawn from a semi-analytical model of the formation and evolution of 
galaxies. We use the predicted average $\left<v_{12}\right>$ to split
the model galaxies into different velocity regimes in order to check the principles upon which the MB test is based.

\section{The galaxy formation model}\label{sec:model}

Galaxies are thought to form within haloes of dark matter, whose gravity allows the galaxies to exist. The formation and evolution of galaxies is affected by a multitude of other processes besides gravity and computational modelling is the only way we can attempt to understand all these processes. For this study we use the {\sc galform} semi-analytical model \citep{cole00}. Semi-analytical models use simple, physically motivated equations to follow the
fate of baryons in a universe in which structure grows hierarchically
through gravitational instability \citep[see][]{baugh06}. In particular, we use the 
\citet{gonzalez13} model, which exploits a Millennium Simulation class N-body run performed with the WMAP7 cosmology \citep{wmap7}: matter density, $\Omega_{\rm m0}=0.272$, cosmological constant,
$\Omega_{\Lambda 0} = 0.728$, baryon density, $\Omega_{\rm b0}=0.045$, a normalisation of density fluctuations given by $\sigma_{8}=0.807$ and \
a Hubble constant
today of $H_0=100\,h$ km$\,{\rm s}^{-1}$Mpc$^{-1}$, with
$h=0.704$. The \citeauthor{gonzalez13} model accounts for the physical 
processes shaping the formation and
evolution of galaxies, including: (i) the collapse and merging of
dark matter haloes; (ii) the shock-heating and radiative cooling of
gas inside dark matter haloes, leading to the formation of galactic
discs; (iii) the quiescent star formation in galactic discs, for which the mass of molecular and atomic gas content is followed explicitly \citep{lagos10}; (iv) feedback
from supernovae, from active galactic nuclei  and from
photoionization of the inter galactic medium; (v) chemical enrichment of the 
stars and gas; (vi) galaxy
mergers driven by dynamical friction within common dark matter haloes,
leading to the formation of stellar spheroids, which also may trigger
bursts of star formation. The end product of the calculation is a prediction 
of the number and properties of galaxies that reside
within dark matter haloes of different masses. The free parameters in the 
\citeauthor{gonzalez13} model were chosen in order to reproduce the rest-frame 
luminosity functions in b$_{J}$ and K-bands at $z=0$ and to give a reasonable 
match to the observed evolution of the rest-frame ultra violet and K-band 
luminosity functions.

\section{Results}
  
\begin{figure}[t!]
 \centering
 \includegraphics[width=0.44\textwidth,clip]{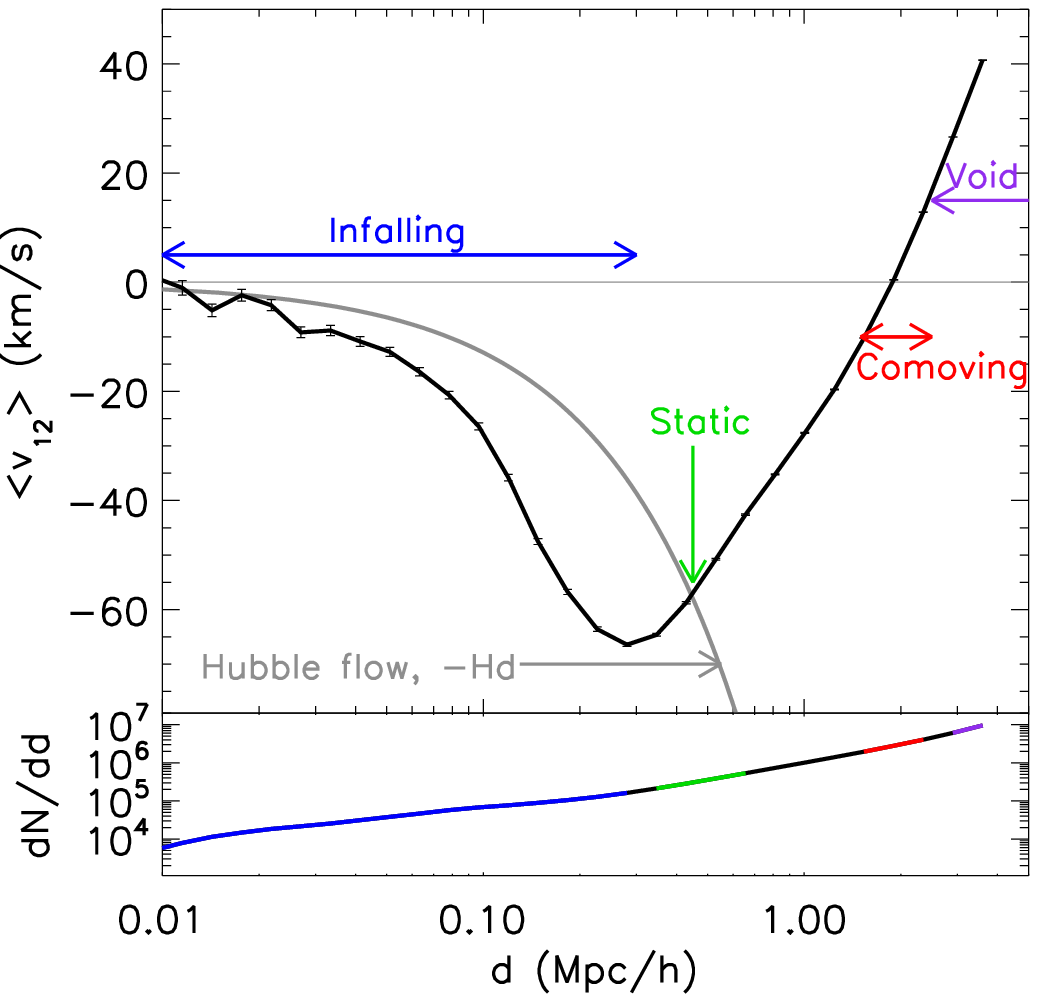}      
 \includegraphics[width=0.52\textwidth,clip]{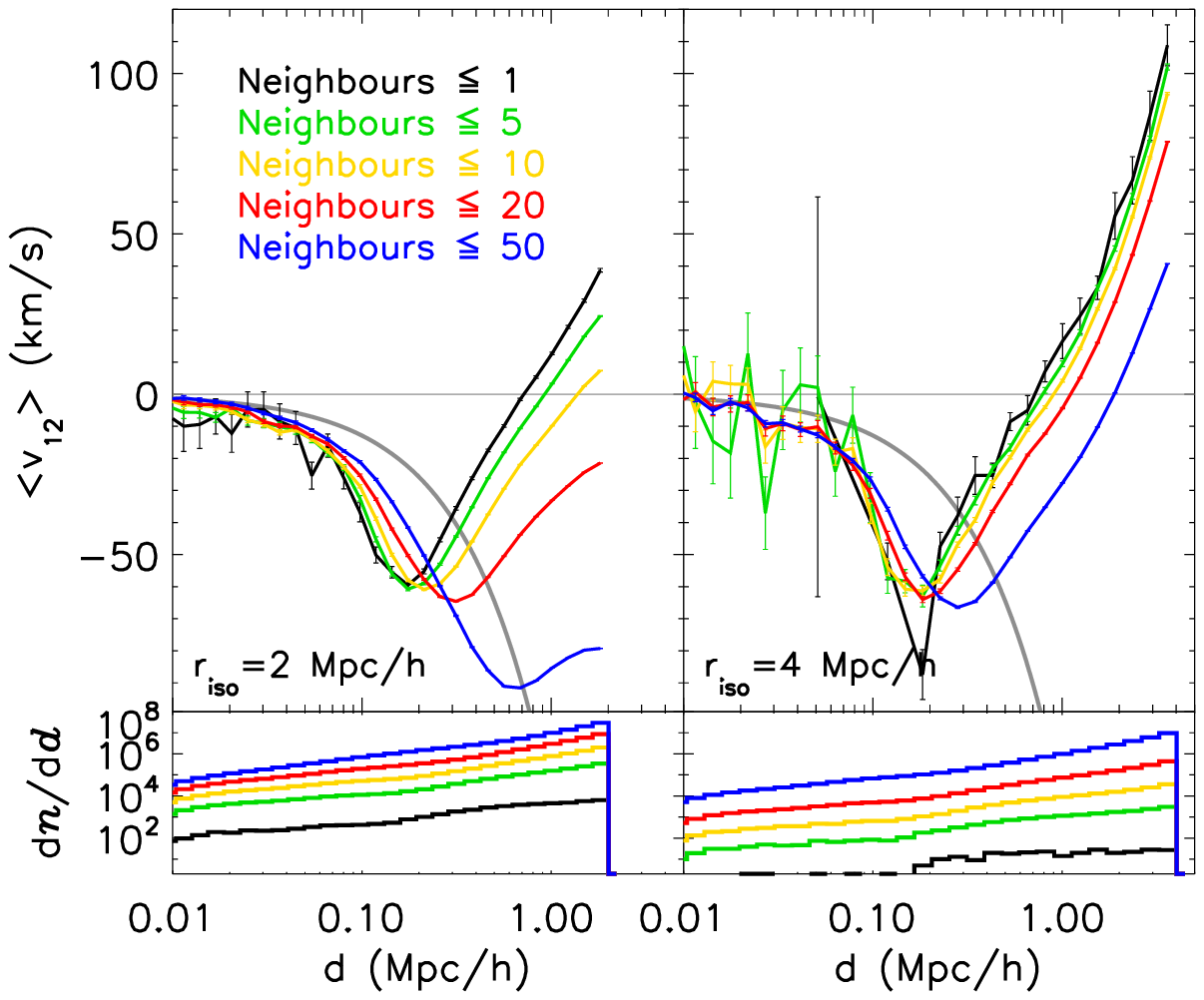}      
  \caption{{\it Left panel:} In the main panel, the black solid line shows the mean comoving relative velocity, $\left<v_{12}\right>$, of pairs of galaxies at $z= 0.5$ as a function of their comoving real-space separation, $d$, as measured from simulations. The pairs of galaxies were selected among those with a maximum of 50 neighbours within a sphere of comoving radius $r_{iso}=4 h^{-1}$Mpc. The different regimes are indicated by arrows. {\it Right panel:} $\left<v_{12}\right>$ as a function of $d$ for pairs of galaxies selected by assuming $r_{iso}=2 h^{-1}$Mpc (left) and $r_{iso}=4 h^{-1}$Mpc (right), and different $N_{max}$, as indicated in the legend. In both panels, the thick grey solid line shows the Hubble flow, $-H(z)d$, and the lower panels show the number of isolated pairs of galaxies per bin in comoving distance.
}
  \label{fig:v12}
\end{figure}

We use model galaxies at $z=0.5$\footnote{In the future, we would like to apply the MB test to galaxies from BOSS, which have $\left<z\right>\sim 0.5$ \citep{boss}.} to study the comoving relative velocities of isolated pairs of galaxies, $v_{12}$. The isolation criteria is defined by counting the number of neighbours a galaxy has within a sphere of a given comoving radius, $r_{iso}\,h^{-1}$Mpc. If that number is below a certain threshold, $N_{max}$, we consider the galaxy to be isolated and, thus, it can become part of an isolated pair of galaxies (note that in this way a galaxy can be part of more than one pair). The LOS radial comoving distance to a galaxy with redshift $z_A$ can be defined as (c is the speed of light):  \begin{equation}
  \chi(z_A)=\frac{c}{H_0}\int_0^{z_{A}} \frac{\mathrm{d}z}{E(z)}\,, \quad {\rm with} \quad E(z)=\sqrt{\Omega_m(1+z)^3+\Omega_k(1+z)^2+\Omega_{\Lambda}}\quad {\rm and} \quad H(z)=H_0E(z).
\end{equation}
The Hubble expansion rate, $H(z)$, can also be defined in terms of the expansion parameter, $a=1/(1+z)$, as $H=\dot{a}/a$. The physical distance to a galaxy, $r$, is proportional to its comoving distance, $\chi$: $r=a\chi$. Thus, if we neglect higher order terms of the total velocity of a galaxy, $v$, we can express $v$ as the sum of the Hubble flow and the peculiar velocity, $v_{p}$: $v=\dot{r}=\dot{a}\chi+a\dot{\chi}=H(z)r+v_{p}$, or in comoving coordinates, $(1+z)v=H(z)\chi+(1+z)v_{p}$. For a pair of galaxies, we can define $v_{12}$ as the rate of change in their comoving real-space separation $\dot{d}$ \citep{bb}:
\begin{equation}
v_{12}=
(1+z)\Delta\overrightarrow{v_p}\hat{d} \, , \quad {\rm with} \quad \Delta\overrightarrow{v_p}=\overrightarrow{v_{p2}}-\overrightarrow{v_{p1}} \quad {\rm (difference\, in\, peculiar\, velocities)} \quad {\rm and } \quad \hat{d} \, {\rm  \, unitary \, vector}
\end{equation}

Thus, when $v_{12}=0$, the pair of galaxies will follow the Hubble flow, $(1+z)(v_2-v_1)=H(z)d$. If $v_{12}=-H(z)d$ the peculiar velocities of the galaxies will be compensating the Hubble flow, i.e. they will be static with respect to each other in physical space. When $v_{12}<-H(z)d$, the galaxies will be infalling and when $v_{12}>-H(z)d$, they will be moving apart. These different regimes are indicated in the left panel in Fig. \ref{fig:v12}, which shows the predicted average $\left<v_{12}\right>$ for isolated model galaxy pairs. The right panel in Fig. \ref{fig:v12} shows the average $\left<v_{12}\right>$ for model galaxy pairs at $z= 0.5$ when different isolation criteria are applied. Increasing $N_{max}$ implies that larger separations between the two galaxies are needed in order for their $\left<v_{12}\right>$ to approach zero. In fact, if no isolation criteria is applied, on average, no galaxy pairs are found with $\left<v_{12}\right>=0$ \citep[see also][]{bb}. Reducing $r_{iso}$ can have a similar effect as increasing $N_{max}$. Applying observational limits to the model galaxies, such as a range in magnitude or stellar mass, has a similar effect to changing the $N_{max}$. As an example, pairs of galaxies with $N_{max}=5$, $r_{iso}=4h^{-1}$Mpc and $m_{AB}(i)<20$ have $\left<v_{12}\right>\sim 0$ only for separations close to 2$h^{-1}$Mpc, instead of the 1$h^{-1}$Mpc. This happens because when we only take into account either bright or massive galaxies, we are missing galaxy neighbours that interact with those galaxies passing the selection criteria and are thus affecting their peculiar velocities. We have also studied the predicted  $\left<v_{12}\right>$ in simulations with a dynamical dark energy \citep{jennings10,jennings12}. We find that $\left<v_{12}\right>$ is less sensitive to changes in the cosmology from an evolving dark energy, than to changes in $N_{max}$.

\begin{figure}[t!]
 \centering
 \includegraphics[width=0.48\textwidth,clip]{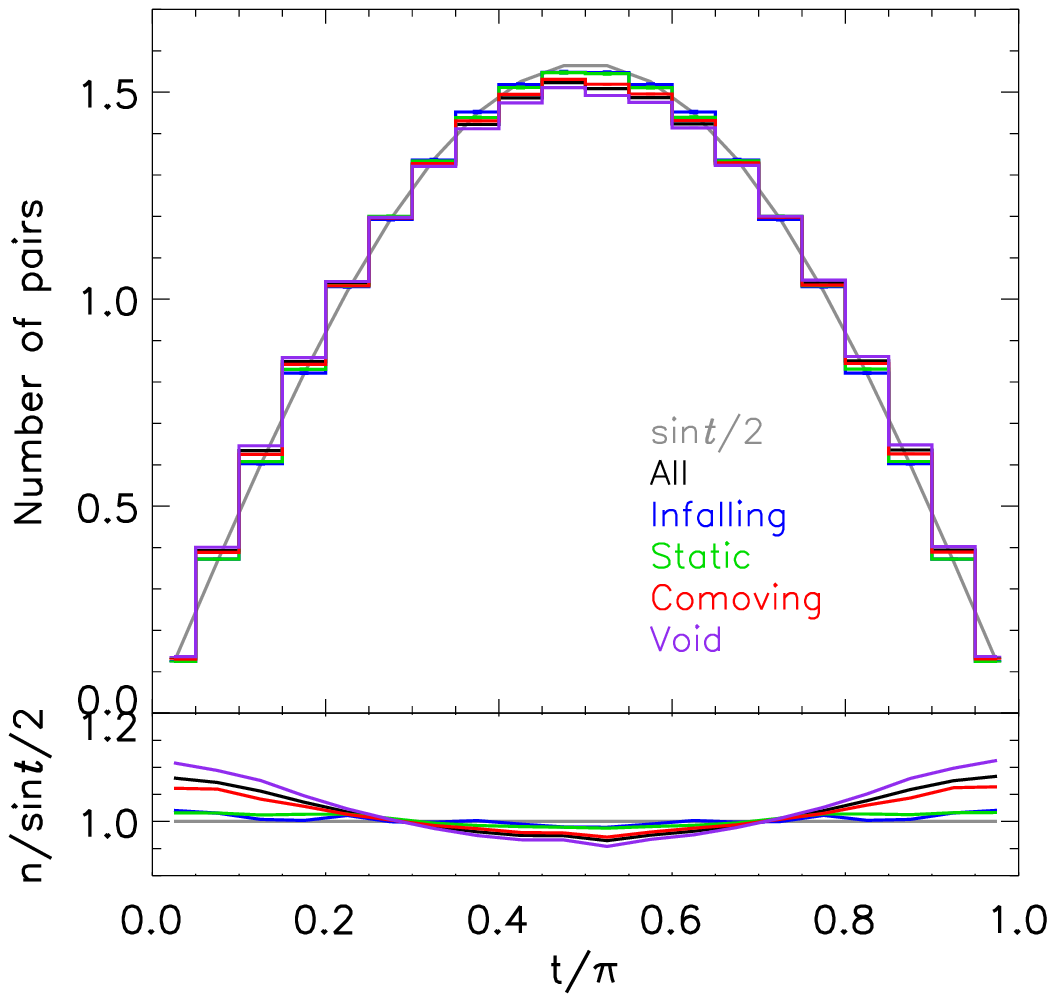}      
 \includegraphics[width=0.48\textwidth,clip]{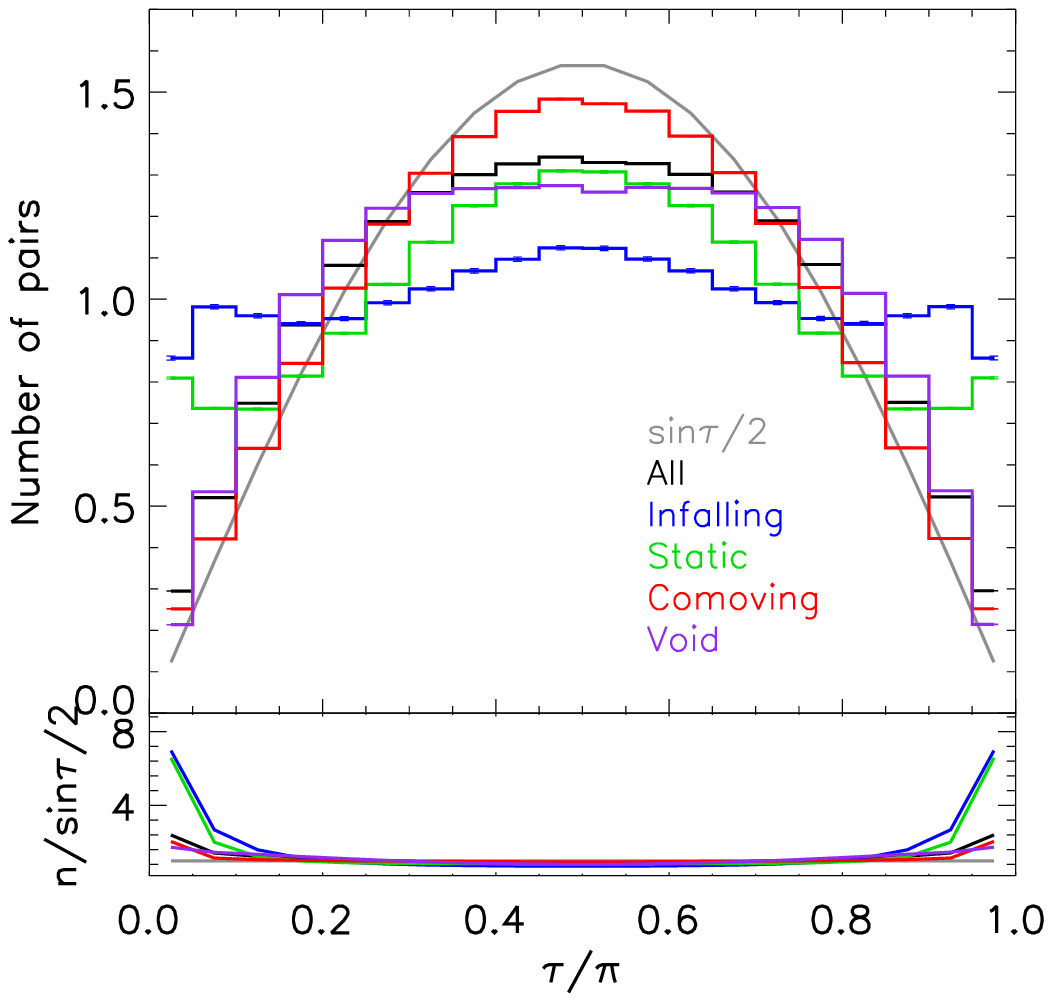}      
  \caption{The main panels show the predicted distribution of the angles that isolated pairs of galaxies form with the LOS in real space ($t$, left) and redshift space ($\tau$, right). All curves are normalized to have unit area. The grey lines show the theoretical expectation of sin $t/2$. The black histograms show the prediction of all modelled isolated pairs of galaxies ($r_{iso}=4\, h^{-1}$Mpc, $N_{max}=50$), the other histograms show the behaviour of the subsample of pairs of galaxies selected by their comoving separation to account for the different regimes of $\left<v_{12}\right>$, as indicated in the legend.  Poisson error bars are shown for each predicted histogram, though they are too small to be seen. The lower panels show the ratio between the predicted distributions and the theoretical expectation.
}
  \label{fig:t}
\end{figure}
\subsection{The distribution of the angle that isolated galaxy pairs form with the line-of-sight}

MB derived a test based on the idea that, in real space, the probability distribution of the angle $t$ between the pair and LOS follows sin $t/2$ with $\left< {\rm sin}^2t\right>=0.667$. We have measured the distribution of the angles $t$ in real space (for a flat universe: sin $t=$sin $\theta\cdot \chi _{z_B}/d$) and $\tau$ in redshift space for the model galaxies. Fig. \ref{fig:t} shows the distribution of these angles for all isolated pairs. In real space, the distribution for isolated model pairs of galaxies is very close to the theoretical expectation of sin $t/2$, with a difference smaller than a factor 1.1. In redshift space, when peculiar velocities are included to estimate the observed positions of galaxies, there is an increase in pairs which are aligned along the LOS and we find deviations from the uniform distribution which are largest for angles $\tau/\pi< 0.2$ and $\tau/\pi>0.8$. 

We have split the isolated pairs of galaxies according to their comoving real-space separations, $d$, which defines the different $\left<v_{12}\right>$ regimes described above. In real space, the distribution of angles are rather insensitive to the cut in $d$. The distributions are at most a factor of 1.2 different with respect to the expected sin $t/2$. However, in redshift space, the distortion of the distribution of $\tau$ does depend on the value of $\left<v_{12}\right>$, since this is a measure of the relative peculiar velocities of the pair. Pairs of galaxies with $\left<v_{12}\right>\sim 0$ show a very small change between the distributions of $t$ in real space and $\tau$ in redshift space. Isolated pairs of galaxies with $N_{max}=50$, $r_{iso}=4h^{-1}$Mpc and in the {\it comoving regime} are predicted to have $\left< {\rm sin}^2t\right>=0.658$ in real space, and $\left< {\rm sin}^2\tau\right>=0.645$ in redshift space, while galaxies in the other regimes are predicted to have a stronger variation of these quantities. Therefore, pairs of galaxies in this regime can be used to constrain cosmology, as we can accurately predict their distribution in redshift space.

\section{Conclusions}
We have used the \citet{gonzalez13} semi-analytical model of galaxy formation and evolution 
to study the average comoving relative velocities of isolated pairs of galaxies, $\left<v_{12}\right>$. This model uses a Millennium Simulation class N-body run performed with the WMAP7 cosmology. We have found that, on average, two isolated galaxies with comoving separations below $\sim 0.5\, h^{-1}$Mpc are infalling towards each other, while two galaxies with separations above $\sim 2\, h^{-1}$Mpc are moving apart. For separations somewhere in between, a regime is found for which $\left<v_{12}\right>\sim 0$. The exact range in comoving real-space distances that define this {\it comoving regime} strongly depends on the definition of the isolation criteria used to select pairs of galaxies. For isolated pairs of model galaxies at $z=0.5$ we have measured in real space the angle that they form with the LOS, finding an excellent agreement with the theoretical expectation for uniformly distributed pairs of galaxies. This finding is practically independent of the average $\left<v_{12}\right>$ and, thus, validates the assumption made by MB for their cosmological test. In redshift space, due to the effect of peculiar velocities, we find more pairs of galaxies which appear closely aligned with the LOS, though the deviations do depend on $\left<v_{12}\right>$. We find that isolated pairs of galaxies selected with $\left<v_{12}\right>\sim 0$ form an angle with the LOS which is practically independent of the effect of peculiar velocities and so are uniformly distributed in both real and redshift space, with $\left< {\rm sin}^2t\right>/\left< {\rm sin}^2\tau\right> \sim 1$. Pairs of galaxies in this regime are therefore a potentially promising probe of cosmology.

\begin{acknowledgements}

{\sc galform} was run on the ICC Cosmology Machine, 
which is part of the DiRAC Facility jointly funded by STFC, the Large Facilities 
Capital Fund of BIS, and Durham University. VGP acknowledges financial
support from the grant OMEGA ANR-11-JS56-003-01.
\end{acknowledgements}

\bibliographystyle{aa}  
\bibliography{gonzalezperez}

\begin{thebibliography}{14}
\expandafter\ifx\csname natexlab\endcsname\relax\def\natexlab#1{#1}\fi

\bibitem[{{Alcock} \& {Paczynski}(1979)}]{ap79}
{Alcock}, C. \& {Paczynski}, B. 1979, \nat, 281, 358

\bibitem[{{Anderson} {et~al.}(2012){Anderson}, {Aubourg}, {Bailey}, {Bizyaev},
  {Blanton}, {Bolton}, {Brinkmann}, {Brownstein}, {Burden}, {Cuesta}, {da
  Costa}, {Dawson}, {de Putter}, {Eisenstein}, {Gunn}, {Guo}, {Hamilton},
  {Harding}, {Ho}, {Honscheid}, {Kazin}, {Kirkby}, {Kneib}, {Labatie},
  {Loomis}, {Lupton}, {Malanushenko}, {Malanushenko}, {Mandelbaum}, {Manera},
  {Maraston}, {McBride}, {Mehta}, {Mena}, {Montesano}, {Muna}, {Nichol},
  {Nuza}, {Olmstead}, {Oravetz}, {Padmanabhan}, {Palanque-Delabrouille}, {Pan},
  {Parejko}, {P{\^a}ris}, {Percival}, {Petitjean}, {Prada}, {Reid}, {Roe},
  {Ross}, {Ross}, {Samushia}, {S{\'a}nchez}, {Schlegel}, {Schneider},
  {Sc{\'o}ccola}, {Seo}, {Sheldon}, {Simmons}, {Skibba}, {Strauss}, {Swanson},
  {Thomas}, {Tinker}, {Tojeiro}, {Maga{\~n}a}, {Verde}, {Wagner}, {Wake},
  {Weaver}, {Weinberg}, {White}, {Xu}, {Y{\`e}che}, {Zehavi}, \&
  {Zhao}}]{anderson12}
{Anderson}, L., {Aubourg}, E., {Bailey}, S., {et~al.} 2012, \mnras, 427, 3435

\bibitem[{{Baugh}(2006)}]{baugh06}
{Baugh}, C.~M. 2006, Reports of Progress in Physics, 69, 3101

\bibitem[{{Blake} {et~al.}(2011){Blake}, {Davis}, {Poole}, {Parkinson},
  {Brough}, {Colless}, {Contreras}, {Couch}, {Croom}, {Drinkwater}, {Forster},
  {Gilbank}, {Gladders}, {Glazebrook}, {Jelliffe}, {Jurek}, {Li}, {Madore},
  {Martin}, {Pimbblet}, {Pracy}, {Sharp}, {Wisnioski}, {Woods}, {Wyder}, \&
  {Yee}}]{blake11}
{Blake}, C., {Davis}, T., {Poole}, G.~B., {et~al.} 2011, \mnras, 415, 2892

\bibitem[{{Bueno Belloso} {et~al.}(2012){Bueno Belloso}, {Pettinari}, {Meures},
  \& {Percival}}]{bb}
{Bueno Belloso}, A., {Pettinari}, G.~W., {Meures}, N., \& {Percival}, W.~J.
  2012, \prd, 86, 023530

\bibitem[{{Cole} {et~al.}(2000){Cole}, {Lacey}, {Baugh}, \& {Frenk}}]{cole00}
{Cole}, S., {Lacey}, C.~G., {Baugh}, C.~M., \& {Frenk}, C.~S. 2000, \mnras,
  319, 168

\bibitem[{{Dawson} {et~al.}(2013){Dawson}, {Schlegel}, {Ahn}, {Anderson},
  {Aubourg}, {Bailey}, {Barkhouser}, {Bautista}, {Beifiori}, {Berlind},
  {Bhardwaj}, {Bizyaev}, {Blake}, {Blanton}, {Blomqvist}, {Bolton}, {Borde},
  {Bovy}, {Brandt}, {Brewington}, {Brinkmann}, {Brown}, {Brownstein}, {Bundy},
  {Busca}, {Carithers}, {Carnero}, {Carr}, {Chen}, {Comparat}, {Connolly},
  {Cope}, {Croft}, {Cuesta}, {da Costa}, {Davenport}, {Delubac}, {de Putter},
  {Dhital}, {Ealet}, {Ebelke}, {Eisenstein}, {Escoffier}, {Fan}, {Filiz Ak},
  {Finley}, {Font-Ribera}, {G{\'e}nova-Santos}, {Gunn}, {Guo}, {Haggard},
  {Hall}, {Hamilton}, {Harris}, {Harris}, {Ho}, {Hogg}, {Holder}, {Honscheid},
  {Huehnerhoff}, {Jordan}, {Jordan}, {Kauffmann}, {Kazin}, {Kirkby}, {Klaene},
  {Kneib}, {Le Goff}, {Lee}, {Long}, {Loomis}, {Lundgren}, {Lupton}, {Maia},
  {Makler}, {Malanushenko}, {Malanushenko}, {Mandelbaum}, {Manera}, {Maraston},
  {Margala}, {Masters}, {McBride}, {McDonald}, {McGreer}, {McMahon}, {Mena},
  {Miralda-Escud{\'e}}, {Montero-Dorta}, {Montesano}, {Muna}, {Myers},
  {Naugle}, {Nichol}, {Noterdaeme}, {Nuza}, {Olmstead}, {Oravetz}, {Oravetz},
  {Owen}, {Padmanabhan}, {Palanque-Delabrouille}, {Pan}, {Parejko},
  {P{\^a}ris}, {Percival}, {P{\'e}rez-Fournon}, {P{\'e}rez-R{\`a}fols},
  {Petitjean}, {Pfaffenberger}, {Pforr}, {Pieri}, {Prada}, {Price-Whelan},
  {Raddick}, {Rebolo}, {Rich}, {Richards}, {Rockosi}, {Roe}, {Ross}, {Ross},
  {Rossi}, {Rubi{\~n}o-Martin}, {Samushia}, {S{\'a}nchez}, {Sayres}, {Schmidt},
  {Schneider}, {Sc{\'o}ccola}, {Seo}, {Shelden}, {Sheldon}, {Shen}, {Shu},
  {Slosar}, {Smee}, {Snedden}, {Stauffer}, {Steele}, {Strauss}, {Streblyanska},
  {Suzuki}, {Swanson}, {Tal}, {Tanaka}, {Thomas}, {Tinker}, {Tojeiro},
  {Tremonti}, {Vargas Maga{\~n}a}, {Verde}, {Viel}, {Wake}, {Watson}, {Weaver},
  {Weinberg}, {Weiner}, {West}, {White}, {Wood-Vasey}, {Yeche}, {Zehavi},
  {Zhao}, \& {Zheng}}]{boss}
{Dawson}, K.~S., {Schlegel}, D.~J., {Ahn}, C.~P., {et~al.} 2013, \aj, 145, 10

\bibitem[{{Gonzalez-Perez} {et~al.}(2013){Gonzalez-Perez}, {Lacey}, {Baugh},
  {Lagos}, {Helly}, \& {Campbell}}]{gonzalez13}
{Gonzalez-Perez}, V., {Lacey}, C.~G., {Baugh}, C.~M., {et~al.} 2013, arXiv
  1309.7057

\bibitem[{{Jennings} {et~al.}(2010){Jennings}, {Baugh}, {Angulo}, \&
  {Pascoli}}]{jennings10}
{Jennings}, E., {Baugh}, C.~M., {Angulo}, R.~E., \& {Pascoli}, S. 2010, \mnras,
  401, 2181

\bibitem[{{Jennings} {et~al.}(2012){Jennings}, {Baugh}, \&
  {Pascoli}}]{jennings12}
{Jennings}, E., {Baugh}, C.~M., \& {Pascoli}, S. 2012, \mnras, 420, 1079

\bibitem[{{Komatsu} {et~al.}(2011){Komatsu}, {Smith}, {Dunkley}, {Bennett},
  {Gold}, {Hinshaw}, {Jarosik}, {Larson}, {Nolta}, {Page}, {Spergel},
  {Halpern}, {Hill}, {Kogut}, {Limon}, {Meyer}, {Odegard}, {Tucker}, {Weiland},
  {Wollack}, \& {Wright}}]{wmap7}
{Komatsu}, E., {Smith}, K.~M., {Dunkley}, J., {et~al.} 2011, \apjs, 192, 18

\bibitem[{{Lagos} {et~al.}(2011){Lagos}, {Lacey}, {Baugh}, {Bower}, \&
  {Benson}}]{lagos10}
{Lagos}, C.~D.~P., {Lacey}, C.~G., {Baugh}, C.~M., {Bower}, R.~G., \& {Benson},
  A.~J. 2011, \mnras, 416, 1566

\bibitem[{{Marinoni} \& {Buzzi}(2010)}]{mb}
{Marinoni}, C. \& {Buzzi}, A. 2010, \nat, 468, 539

\bibitem[{{Phillipps}(1994)}]{phillipps94}
{Phillipps}, S. 1994, \mnras, 269, 1077

\end{thebibliography}

\end{document}